\documentclass[12pt]{article}
\usepackage{graphicx}
\def\la{\mathrel{\mathpalette\fun <}}
\def\ga{\mathrel{\mathpalette\fun >}}
\def\fun#1#2{\lower3.6pt\vbox{\baselineskip0pt\lineskip.9pt
  \ialign{$\mathsurround=0pt#1\hfil##\hfil$\crcr#2\crcr\sim\crcr}}}
\begin{document}

\title{Ultra-High Energy Cosmic Rays: A Probe of Physics and
Astrophysics at Extreme Energies}

\author{G{\"u}nter Sigl\\
Institut d'Astrophysique de Paris, 98bis Boulevard Arago, CNRS\\
75014 Paris, France}

\maketitle

The origin of cosmic rays is one of the major unresolved questions
in astrophysics. In particular, the highest energy cosmic
rays observed possess macroscopic energies and their origin is
likely associated with the most energetic processes in the Universe.
They thus provide a probe of physics and astrophysics at energies
that are unreached in laboratory experiments. Theoretical explanations
range from acceleration of charged particles in astrophysical
environments to particle physics beyond the well established Standard Model,
and processes taking place at the earliest moments of our Universe.
Distinguishing between these scenarios requires detectors with
effective areas in the 1000 km$^2$ range which are now under
construction or in the planning stage. Close connections with
gamma-ray and neutrino astrophysics add to the interdisciplinary
character of this field.

\newpage

High energy cosmic ray (CR) particles are shielded
by the Earth's atmosphere and reveal their existence on the
ground only by indirect effects such as ionization and
showers of secondary charged particles covering areas up
to many km$^2$ for the highest energy particles. In fact,
in 1912 Victor Hess discovered CRs by measuring ionization from
a balloon~\cite{hess}, and in 1938 Pierre Auger proved the existence of
``extensive air showers'' (EAS) caused by primary particles
with energies above $10^{15}\,$eV by simultaneously observing
the arrival of secondary particles in ground detectors many meters
apart~\cite{auger_disc}.

After almost 90 years of CR research, their origin
is still an open question, with a degree of uncertainty
increasing with energy~\cite{crbook}: Only below 100 MeV
kinetic energy, where the solar wind shields protons coming
from outside the solar system, the Sun must give rise to
the observed proton flux. Above that energy the CR spectrum
exhibits surprisingly little structure and is well approximated
by broken power laws $\propto E^{-\gamma}$ (see Fig.~\ref{fig1}):
At the energy $E\simeq4\times 10^{15}\,$eV
called the ``knee'', the flux of particles per area, time, solid angle,
and energy steepens from a power law index $\gamma\simeq2.7$
to one of index $\simeq3.0$. The bulk of the CRs up to at least
that energy is believed to originate within our Galaxy. Above the so
called ``ankle'' at $E\simeq5\times10^{18}\,$eV, the spectrum flattens
again to a power law of index $\gamma\simeq2.8$. This latter feature
is often interpreted as a cross over from a steeper Galactic
component, which above the ankle cannot be confined by the Galactic
magnetic field, to a harder component
of extragalactic origin. At the highest energies there is no
apparent end to the CR spectrum, and over the last few years
giant air showers from CR primaries with energies exceeding
$10^{20}\,$eV~\cite{data1,data} (see Fig.~\ref{fig2}) have
been detected. This represents up to 50
Joules in what appears to be one elementary particle, about $10^8$ times
higher than energies achievable in man made accelerators! The nature and
origin of CRs above the ankle, which we will call ultra-high
energy cosmic rays (UHECRs), and especially the ones above
$10^{20}\,$eV are mysterious~\cite{bs,reviews} and will be the main
focus of this review.

The conventional ``bottom-up'' scenario assumes that all high energy
charged particles are accelerated in astrophysical environments,
typically in magnetized astrophysical shocks. A general estimate of
the maximal energy that can be achieved
is given by the requirement that the gyroradius $r_g\simeq E/(ZeB)$ of
the particle of charge $Ze$ and energy $E$ in a magnetic field
$B$ is smaller than the size $R$ of the accelerator, in numbers
\begin{equation}
  r_g\simeq100\,Z^{-1}\left(\frac{E}{10^{20}\,{\rm eV}}\right)
  \left(\frac{B}{\mu{\rm G}}\right)^{-1}\,{\rm kpc};\quad
  E\la10^{18}\,Z\left(\frac{R}{{\rm kpc}}\right)
  \left(\frac{B}{\mu{\rm G}}\right)\,{\rm eV}\,.\label{Emax}
\end{equation}
Here, $B$ is measured in micro Gauss ($\mu$G) and $R$ in kilo parsec
($1\,{\rm pc}=3.09\times10^{18}\,$cm). Eq.~(\ref{Emax})
is an optimistic estimate since it neglects the finite
lifetime of the accelerator and energy losses due to interactions
with the ambient environment such as synchrotron radiation
in the magnetic field and production of secondary particles.
Apart from the different scales, the factors involved are the same
ones which limit the maximal energy in accelerator laboratories.
The remnants associated with Galactic supernova explosions have
sizes up to $R\sim\,$pc with magnetic fields up to the milli Gauss
range. According to Eq.~(\ref{Emax}) they should thus be able to
accelerate CRs at least up to the knee, possibly up to the ankle.
This and the fact that the power required to maintain the cosmic
ray density in our Galaxy is comparable to the kinetic energy output
rate of Galactic supernovae suggests that supernovae are the predominant
sources of CRs in this energy
range. At higher energies powerful extragalactic objects
such as active galactic nuclei (AGN) are envisaged~\cite{crbook}.
However, the existence of UHECRs at energies around $10^{20}\,$eV
and above, assuming them
to be one of the known electromagnetically or strongly interacting
particles, poses at least three theoretical problems:

\section*{Extragalactic Sources and the ``GZK Cutoff''}
Interactions with the omnipresent 2.7 Kelvin cosmic microwave
background radiation (CMB), which is a thermal relic of the big bang,
limit the attenuation length of the highest energy particles to less
than about 50 megaparsecs (Mpc).
For example, in the rest frame of a nucleon of energy $E\ga E_{th}$
the CMB will appear as a background of $\gamma-$rays of sufficiently
high energy to allow the production of pions. The threshold energy
is given by
\begin{equation}
  E_{th}=\frac{m_\pi(m_N+m_\pi/2)}{\varepsilon}\simeq
  6.8\times10^{19}\left(\frac{\varepsilon}
  {10^{-3}\,{\rm eV}}\right)^{-1}\,{\rm eV}\,,\label{pionprod}
\end{equation}
where $m_N$ and $m_\pi$ are the nucleon and pion mass,
respectively, and $\varepsilon\sim10^{-3}\,$eV is a typical
CMB photon energy. For $E\ga E_{th}$ the nucleon will loose a significant
part of its energy on a length scale of $l_\pi\simeq1/(\sigma_\pi n_{\rm CMB})
\simeq20\,$Mpc, where $n_{\rm CMB}\simeq422\,{\rm cm}^{-3}$ is the
number density of CMB photons, and the pion production cross
section $\sigma_\pi\sim100\,{\rm mbarn}=10^{-25}\,{\rm cm}^2$.
Nuclei and $\gamma-$rays have similar energy loss distances due to
photodisintegration and electron-positron pair
production on the CMB, respectively~\cite{bs}. Therefore, if
the CR sources were all at cosmological distances
(i.e. several thousand Mpc away), the energy spectrum would exhibit
a depletion of particles above a few $10^{19}\,$eV, the
so-called Greisen-Zatsepin-Kuzmin (GZK)\cite{gzk} cutoff.
Since the data do not confirm such a cutoff~\cite{data1,data}
(see Fig.~\ref{fig2}), an astrophysical origin would require the
sources to be within about 100 Mpc. The only way to avoid
this conclusion without invoking as yet unknown new physics is that charged
particles accelerated in sources at much larger distances give
rise to a secondary neutrino beam which can propagate
unattenuated. This neutrino beam has to be sufficiently strong
to produce the observed UHECRs within 100 Mpc by electroweak (EW)
interactions with the relic neutrino background, the neutrino
analogue of the CMB~\cite{zburst1}. However, this requires
extremely powerful sources and local relic neutrino overdensities
that hardly seem consistent with commonly accepted ideas about
cosmic structure formation~\cite{zburst2}. In addition, to
avoid excessive fluxes at lower energies, the sources have to be
nearly opaque to $\gamma-$rays and nucleons~\cite{zburst2}.

\section*{The Maximal Acceleration Energy Problem}
Evaluating the maximum energy estimates in Eq.~(\ref{Emax})
for known astrophysical objects demonstrates that only very few such
objects seem capable of
accelerating charged particles up to a few $10^{20}\,$eV~\cite{acc}.
In our Galactic neighbourhood, pulsars with magnetic fields larger
than $10^{12}\,$G satisfy the criterion Eq.~(\ref{Emax}) for
iron nuclei. But it remains to be seen if energy losses in the
dense pulsar environment do not considerably decrease the maximum
energy~\cite{beo}. Another interesting but highly speculative 
suggestion is the acceleration of particles to such energies in
ultrarelativistic jets from bipolar supernovae in our Galaxy~\cite{dp}.
In general, Galactic sources tend to predict UHECR
arrival directions correlated with Galactic structures, which is not seen
in the data (see below). Possible extragalactic accelerators include
AGN, radio galaxies~\cite{bier-rev}, shock waves associated with
large scale structure formation~\cite{shocks}, and possibly $\gamma-$ray
bursts. AGN are numerous enough, but are unlikely to reach the
requisite energies, due to strong energy losses in the intense
radiation fields of the cores. Hot spots in the jets of radio
galaxies are sufficiently tenuous to avoid excessive energy losses,
and extend up to kpc scales. With fields in the milli Gauss regime
they meet the requirement Eq.~(\ref{Emax}) and synchrotron observations
even seem to require the presence of protons up to
$\sim10^{21}\,$eV in these objects~\cite{radio}.
The main problem is that such objects are rare~\cite{bier-rev}.
Gamma-ray bursts, another as yet not understood enigma of astrophysics,
have been observed to occur with a rate of about one burst
within 100 Mpc (the maximal source distance for nucleons) per 100 years,
each emitting up to $\sim10^{54}\,$ergs in $\gamma-$rays (depending
on the unknown amount of beaming) within a
few seconds. Therefore, if $\gamma-$ray bursts are to explain
the few dozens of UHECRs observed within the past few decades above
the GZK cutoff, they have to meet the following requirements:
They must emit at least as much energy in the form of UHECRs as
in $\gamma-$rays in the MeV range~\cite{stecker}, the UHECRs
must be charged, and their arrival times must be spread out by at
least a few hundred years. The latter requires large scale magnetic
fields stronger than about $10^{-10}\,$G on Mpc scales~\cite{waxman}.

\section*{Angular Distributions and Missing Counterparts}
The seeming isotropy on large angular scales (with a
possibly significant interesting clustering on degree scales) of
arrival directions up to the highest energies~\cite{agasa_clu} leaves
only two possibilities for the source locations: Either there must be many
nearby sources, at least one close to each arrival direction.
Sufficiently powerful astrophysical accelerators which meet the
above criteria are rare and should easily be detected within 100 Mpc,
but no convincing source candidates
have been found~\cite{sources}. Or, alternatively, there are
only very few nearby sources which then requires strong
deflection in Galactic and/or extragalactic magnetic
fields within a few Mpc propagation length. Eq.~(\ref{Emax})
shows that this requires fields of at least $\sim10^{-7}\,$G
on Mpc scales. Such high field strengths are indeed expected to
be localized in sheets and clusters of galaxies, but are
hard to measure directly~\cite{strucmag}.
These values are also close to upper limits established from
independent observations such as the frequency dependent
Faraday rotation of the polarization of radio emission from distant
sources in intervening magnetic fields~\cite{mag}.

Whether the expected distribution and strength of magnetic fields
associated with large scale galaxy structure are consistent with
UHECR spectra and angular distributions is currently under
investigation~\cite{highmag1}. As an example~\cite{highmag2},
Fig.~\ref{fig3} shows predictions for the distribution of arrival
times and energies, the sky averaged spectrum, and the angular distribution of
arrival directions in Galactic coordinates. In this scenario
the UHECR sources are continuously distributed according to the matter
density in the Local Supercluster, following an idealized pancake
profile with scale height of 5 Mpc and scale length 20 Mpc, with
no sources within 2 Mpc from the observer. All sources inject an
$E^{-2.4}$ proton spectrum up to $10^{22}\,$eV.
The square of the magnetic field has a Kolmogorov spectrum with a
maximal field strength $B_{\rm max}=5\times10^{-7}\,$G in the plane
center, and also follows the matter density. The observer is within
2 Mpc of the Supergalactic Plane whose location is
indicated by the solid line in the lower panel and at a distance
$d=20\,$Mpc from the plane center.
This example demonstrates the two major points of scenarios
with large scale fields up to a micro Gauss: First, a steepening of the UHECR
spectrum in the diffusive regime {\it below} $\sim10^{20}\,$eV
may help to explain the observed spectrum at least down to $10^{19}\,$eV
with only one source component~\cite{highmag3}. It is not clear, however, if
the predicted flux is high enough above $10^{20}\,$eV. Second, the predicted
sky distribution may still not be isotropic enough unless the sources
are not strongly correlated with the large scale galaxy structure.

Generally, magnetic fields down to $\sim10^{-11}\,$G can leave observable
imprints on UHECR arrival time, energy and direction
distributions~\cite{bs,deflec}.
This may also help to make progress in the question of
``magnetogenesis'', the origin of galactic and cosmological
magnetic fields, which likely have been seeded in the early
Universe~\cite{gr}.

The enigma of UHECR origin is in a certain way opposite to the
dark matter problem: Dark matter is expected to exist because
of cosmological reasons~\cite{kt} but has not been found
yet, whereas UHECRs above the GZK cutoff were not expected
to exist but are convincingly observed! In recent years this
challenge triggered many theoretical proposals
for the origin of these highest energy particles in the Universe,
as well as new experimental projects and the study of new
detection concepts. We first summarize the experimental
activities.

\section*{Pioneering Experiments and New Detection Concepts}
Above $\sim10^{14}\,$eV, the showers of secondary particles
created by interactions of the primary CRs
in the atmosphere are extensive enough to be detectable from
the ground. In the most traditional technique, charged hadronic
particles, as well as electrons and muons in these
EAS are recorded by detecting the Cherenkov light that they
emit when passing through water tanks,
or by using scintillation counters. Apart from earlier experiments
that were operative between the 1960s and 1980s~\cite{data}, this technique is
used by the largest operating ground array, the Akeno
Giant Air Shower Array (AGASA) near Tokyo, Japan,
covering an area of roughly $100\,{\rm km}^2$
with about 100 detectors of a few meters in size, mutually separated
by about $1\,$km~\cite{agasa}.
Given a flux of about one particle per km$^2$ per century above
$10^{20}\,$eV (see Fig.~\ref{fig2}), the detection rate for
such particles is less than one per year with such an instrument.
The ground array technique allows one to
measure a lateral cross section of the shower profile and
to estimate the energy of the shower-initiating primary particle
from the density of secondary charged particles.

EAS can also be detected via the virtually isotropic fluorescence
emission of the air nitrogen that they excite.
A system of mirrors and photomultipliers in the form of
an insect's eye can be used to track the longitudinal development
of EAS. This technique was first used by the Fly's Eye detector~\cite{fe}
and will be part of several future projects on UHECR detection
(see below). The primary energy can be estimated from
the total fluorescence yield and the longitudinal shower
shape contains information about the primary
composition. Comparison of CR spectra measured with the ground array
and the fluorescence technique indicate systematic errors in
energy calibration that are generally smaller than $\sim$ 40\%~\cite{data}.

An upscaled version of the old Fly's Eye experiment, the
High Resolution Fly's Eye detector already takes data
in Utah, USA~\cite{hires}. Taking into account a duty cycle of about
10\% (a fluorescence detector requires clear, moonless nights),
this instrument will collect events above $10^{17}\,$eV at a rate 
about 10 times larger than for the old Fly's Eye.
Another project utilizing the fluorescence technique
is the Japanese Telescope Array~\cite{tel_array} which is currently
in the proposal stage. If approved, its collecting power will also
be about 10 times that of the old Fly's Eye above $10^{17}\,$eV.
The largest project presently under construction is the  Pierre Auger
Giant Array Observatory~\cite{auger} planned for two sites, one in
Mendoza, Argentina and another in Utah, USA for maximal sky coverage.
Each site will have  a $3000\,{\rm km}^2$ ground array. The southern
site will have about 1600 particle detectors (separated by 1.5 km
each) overlooked by four fluorescence detectors. The ground arrays
will have a duty cycle of nearly 100\%,
leading to detection rates about 30 times as large as for the AGASA
array, i.e. about 50 events per year above $10^{20}\,$eV. About 10\% 
of the events will be detected by both the ground array
and the fluorescence component and can be used for cross
calibration and detailed EAS studies. The detection energy threshold will
be around $10^{18}\,$eV.

An old idea envisages to detect EAS in the Earth's atmosphere from
space. This would provide an increase by another factor $\sim50$ in
collecting power compared to the Pierre Auger Project, i.e. an
event rate above $10^{20}\,$eV of up to a few thousand per year.
Two concepts are currently being studied, the Orbiting Wide-angle
Light-collector (OWL)~\cite{owl} in the USA and the Extreme Universe
Space Observatory (EUSO)~\cite{euso} in Europe of which a prototype may
fly on the International Space Station.

Space-based detectors would be especially suitable for detection
of very small event rates such as those caused by neutrino
primaries which rarely interact in the atmosphere due to their
small interaction cross sections. This disadvantage for the detection
process is at the same time a blessing since it makes these elusive
particles reach us unattenuated over cosmological distances and from
very dense environments where all other particles (except gravitational
waves) would be absorbed. Giving rise to showers typically starting 
deep within the atmosphere, they are also easy to distinguish
from other primaries. Besides detection from space, several other
concepts are currently under study. These include detection of near-horizontal
air showers with ground arrays~\cite{has}, and detection of radio pulses
emitted by neutrino induced electromagnetic showers within large effective
volumes (see Ref.~\cite{bs} for more details).

All these experimental concepts aim at probing existing theoretical
concepts on the yet unknown origin of the highest-energy particles in the
Universe or discovering new physics at energies unreachable in the
laboratory. Let us now give an idea of what may be in store in
terms of new physics.

\section*{Relics from the Early Universe}
The apparent difficulties of bottom-up acceleration scenarios discussed
earlier motivated the proposal of the ``top-down'' scenarios, where UHECRs,
instead of being accelerated,
are the decay products of certain sufficiently massive ``X'' particles
produced by physical processes in the early Universe.
Furthermore, particle accelerator
experiments and the mathematical structure of the Standard Model
of the weak, electromagnetic and strong interactions suggest that
these forces should be unified at energies of about $2\times10^{16}\,$GeV
($1\,$GeV$=10^9\,$eV)~\cite{gut}, 4-5 orders of magnitude above the
highest energies observed in CRs. The relevant ``Grand Unified Theories''
(GUTs) predict the existence of X particles with mass $m_X$ around this
GUT scale. If their lifetime is comparable or larger than
the age of the Universe, they would be dark matter candidates
and their decays could contribute to UHECR fluxes today, with
an anisotropy pattern that reflects the expected dark matter
distribution. However, in many GUTs supermassive particles are expected
to be very short lived and thus have to be produced continuously if
their decays are to give rise to UHECRs. This can only occur
by emission from topological defects which are relics of cosmological
phase transitions that could have occurred in the early Universe at
temperatures close to the GUT scale. Topological defects
necessarily occur between regions that are causally disconnected, such
that the orientation of the order parameter
associated with the phase transition can not be communicated
between these regions and thus will adopt different
values. Examples are cosmic strings (similar to vortices in superfluid
helium), magnetic monopoles, and domain walls (similar to Bloch
walls separating regions of different magnetization in a ferromagnet).
The defect density is consequently given by the particle horizon
in the early Universe and their formation can by analogy even be studied
in solid state experiments where the expansion rate of the Universe
corresponds to the quenching speed that is applied to induce the
transition~\cite{vachaspati}. The defects are
topologically stable, but in the case of GUTs time dependent motion
can lead to the emission of GUT scale X particles.

It is interesting to note that one of the prime cosmological motivations
to postulate inflation, a phase of exponential expansion in the early
Universe~\cite{kt}, was to dilute excessive production
of ``dangerous relics'' such as topological defects and
superheavy stable particles. However, right after inflation,
when the Universe reheats, phase transitions can occur and such
relics can be produced in cosmologically interesting abundances,
and with a mass scale roughly given by the inflationary scale.
This scale is fixed by the CMB anisotropies to
$\sim10^{13}\,$GeV~\cite{kuz-tak}. The reader will notice that
this mass scale is not far above the highest energies observed in CRs,
thus motivating a connection between these primordial relics and
UHECRs which in turn may provide a probe of the early Universe!

Within GUTs the X particles typically decay into jets of particles
whose spectra can be well estimated within the Standard Model.
Before reaching Earth, the injected spectra are reprocessed
by interactions with the low energy photon backgrounds such as
the CMB, and magnetic fields present in the Universe (see Ref.~\cite{bs,slby}
for details). Fig.~\ref{fig4} shows a typical example for the UHECR
spectrum expected in top-down scenarios: The observed flux
is reproduced above $3\times10^{19}\,$eV; at lower energies where
the Universe is transparent to nucleons, bottom-up mechanisms could
explain the spectrum without significant problems. The X particle sources
are not necessarily expected to be associated with astrophysical
objects, but their distribution has to be sufficiently continuous
to be consistent with observed angular distributions.

The most characteristic features are visible in Fig.~\ref{fig4}:
Electromagnetic cascades induced by interactions of the injected
particles with the
low energy photon backgrounds lead to a strong contribution to
the diffuse $\gamma-$ray flux between 30 MeV and 100 GeV, close
to the flux measured by the EGRET detector flown on board the
Compton $\gamma-$ray Observatory satellite~\cite{egret}.
The energy content in these $\gamma-$rays is comparable to the
one in the ultra-high energy neutrino flux which should be detectable
with next generation experiments (see Fig.~\ref{fig4}). The
neutrino flux is hardly influenced by subsequent interactions,
and thus directly represents the decay spectrum. In bottom-up
scenarios neutrinos can only be produced as secondaries and for
sources transparent to the primary nucleons the neutrino flux
must be considerably smaller~\cite{wbbound}. This can also serve as a
discriminator between the top-down and bottom-up concepts.
Finally, top-down models predict a significant $\gamma-$ray component
above $\sim10^{20}\,$eV, whereas nucleons would dominate at lower energies.
This will be a strong discriminator as experiments will improve
constraints on UHECR composition which currently seem to favor
nucleons~\cite{gamma}.

Besides some uncertainties in the shape and chemical composition
of the spectrum, possibly the most significant shortcoming of top-down
scenarios is their lack of predictivity concerning the absolute
flux normalization. At least, the moderate rate of 10 decays per year in
a spherical volume with radius equal the Earth-Sun distance, the
rate necessary to explain the UHECR flux, is not in a remote
corner of parameter space for most scenarios: Dimensional
and scaling arguments imply that topological defects release
X particles with an average rate at cosmic time $t$ of
\begin{equation}
  \dot n_X(t)=\kappa\,m_X^p\,t^{-4+p}\,,\label{dotnx}
\end{equation}
where the dimensionless parameters $\kappa$ and $p$ depend on the
specific top-down scenario~\cite{bs}. For example, hybrid
defects involving cosmic strings have $p=1$ and normalization
of predicted spectra both at EGRET energies and around
$10^{20}\,$eV, as shown in Fig.~\ref{fig4}, leads to
$\kappa m_X\sim10^{13}-10^{14}\,$GeV. For $\kappa\sim1$,
the resulting mass scale is again close to the inflation
and GUT scales!

\section*{New Primary Particles and New Interactions}
A possible way around the problem of missing counterparts
in the framework of acceleration scenarios is to propose primary
particles whose range is not limited by interactions with
the CMB. The only established candidate is the neutrino.
More speculatively, one could propose as yet undiscovered
neutral particles which, according to Eq.~(\ref{pionprod}),
would have a higher GZK threshold if they are more massive
than nucleons. In fact, in supersymmetric extensions of the
Standard Model, new neutral hadronic bound states of light
gluinos with quarks and gluons, so-called R-hadrons with masses
in the 10 GeV range, have been suggested~\cite{cfk}. However,
this possibility seems difficult to reconcile with accelerator
constraints~\cite{gluino}. Magnetic monopoles and their bound
states~\cite{monopole} as well as superconducting string loops~\cite{bp}
similarly have the advantage of not being degraded significantly
by interactions with the CMB and can be efficiently
accelerated. The main problematic issues with these primaries are
the spectra, the atmospheric shower profiles, and (for
non-relativistic monopoles) the arrival direction distributions.
For example, the latter should show correlations with Galactic
structures which are not observed. We will therefore here
mostly focus on neutrinos.

To rescue the bottom-up scenario, the particle propagating over
extragalactic distances, be it a neutrino or a new massive neutral
hadron has to be produced in
interactions of a charged primary which is accelerated in a powerful
astrophysical object. In comparison to EAS induced by nucleons, nuclei,
or $\gamma-$rays, the accelerator can now be located at cosmological
distances. The cost of this conceptual advantage is an increase of
the necessary charged primary energy to $\ga 10^{22}\,$eV due to
losses caused by the expansion of the Universe (redshift) and in the
production of the secondary. These scenarios predict a correlation
between UHECR arrival directions and sources at cosmological
distances. Possible evidence for an angular correlation of
events above the GZK cutoff with compact radio quasars at several
thousand Mpc distance is currently being debated~\cite{corr}.
Only a few more events could settle the question!

Neutrino primaries have the advantage of being established
particles. Unfortunately, within the
Standard Model their interaction cross section with nucleons,
$\sigma_{\nu N}$, falls short of producing ordinary air showers
by about five orders of magnitude, with significant
ramifications for their detection, as mentioned above.
However, at (squared) center of mass (CM) energies $s$
above the EW scale, corresponding to
$\simeq10^{15}\,$eV in the nucleon rest frame,
this cross section has not been measured. Field theory constraints
on the growth at higher energies based on unitarity are relatively
weak~\cite{gw}. Neutrino induced air showers above $10^{15}\,$eV
may therefore rather directly probe new physics beyond the EW
scale, if it leads to enhanced cross sections.

One theoretical possibility consists of a large increase
in the number of degrees of freedom above the EW 
scale~\cite{kovesi-domokos}. A specific implementation
of this idea is provided by scenarios with additional large
compact dimensions and a string or quantum gravity scale
$M_s\sim\,$TeV ($=10^{12}\,$eV). This concept has recently received much
attention in the literature~\cite{tev-qg} because it may imply
unification of all forces in the TeV range, not far above the scale
of EW interactions. This scenario would avoid the
``hierarchy problem'' between the EW scale $\simeq100\,$GeV
and the Planck scale $\simeq10^{19}\,$GeV of gravity.
The cross sections within such scenarios have
not been calculated from first principles yet, but several
arguments based on unitarity lead to estimates that can
very roughly be parametrized by~\cite{nucross}
\begin{equation}
  \sigma_{\rm new}\simeq\frac{4\pi s}{M^4_s}\simeq
  10^{-27}\left(\frac{M_s}{{\rm TeV}}\right)^{-4}
  \left(\frac{E}{10^{20}\,{\rm eV}}\right)\,{\rm cm}^2\,.
  \label{sigma_graviton}
\end{equation}
In the last expression we specified to a neutrino
of energy $E$ hitting a nucleon at rest. A neutrino
would typically start to interact in the atmosphere
for $\sigma_{\nu N}\ga10^{-27}\,{\rm cm}^2$, i.e. in the
case of Eq.~(\ref{sigma_graviton}) for
$E\ga10^{20}\,$eV, assuming $M_s\simeq1\,$TeV,
a value consistent with lower limits from accelerator
experiments~\cite{cpp} and astrophysical constraints~\cite{sn87a}.
The neutrino therefore becomes a primary candidate for the
observed UHECR events. Cross sections of the form
Eq.~(\ref{sigma_graviton}) would predict the average atmospheric
column depth of the first interaction point of neutrino induced
EAS to depend linearly on energy. This signature should easily
be distinguishable from the logarithmic scaling expected
for nucleons, nuclei, and $\gamma-$rays.

Independent of theoretical arguments, the UHECR data
can be used to put constraints on neutrino cross sections
at energies not accessible in the laboratory:
The Fly's Eye experiment has not seen any air showers developing
deep in the atmosphere and has put a limit on their
rate~\cite{baltrusaitis} (see Fig.~\ref{fig4}).
The existence of a secondary neutrino flux from
the decay of pions produced in UHECR interactions with the CMB
(marked ``$N\gamma$ in Fig.~\ref{fig4}) then
implies that $\sigma_{\nu N}$ cannot be larger than the
Standard Model cross section by more than a factor $\sim10^3$
between $10^{18}\,$eV and $10^{20}\,$eV~\cite{tos}. This
conclusion can only be avoided if UHECRs do not have an
extragalactic origin or if $\sigma_{\nu N}$ is comparable
to hadronic cross sections, giving rise to normal EAS.
The projected sensitivity of future experiments such as
the Pierre Auger Observatories and the space based satellite
projects (see Fig.~\ref{fig4}) indicate that these cross section
limits could be improved by up to four orders of magnitude.

Probably the most radical proposition concerns a violation
of one of the basic symmetry principles of modern field
theory such as Lorentz invariance. Such violations can
kinematically prevent energy loss processes such as pion
production at high Lorentz factors~\cite{lorentz}. A reliable
experimental determination
of source distances and primary composition could confirm such
symmetry violations or constrain them possibly more strongly
than accelerator experiments~\cite{abgg}.

\section*{Conclusions}
UHECRs attest to perhaps the most energetic
processes in the Universe. They are not only messengers of astrophysics
at extreme energies, but may also open a window to particle physics
beyond the Standard Model as well as probing processes
occurring in the early Universe at energies close
to the GUT scale. Furthermore, complementary to other methods
such as Faraday rotation measurements, UHECRs
can be used to probe the poorly known large
scale cosmic magnetic fields and their origin. There seems to be
no single convincing theoretical model for the UHECR origin yet
and thus the solution to this problem will strongly depend on
detailed measurements of energy distributions, arrival
directions and times, and composition. The scientific community
eagerly awaits the arrival of several large scale experiments
under construction or in the proposal stage.

\newpage

\section*{Figure Captions}

Fig.~1: {The CR all-particle spectrum observed
by different experiments above $10^{11}\,$eV (from Ref~\cite{data1}
with permission). The differential flux in units of events per area, time,
energy, and solid angle was multiplied with $E^3$
to project out the steeply falling character. The ``knee''
can be seen at $E\simeq4\times 10^{15}\,$eV, and the ``ankle''
at $E\simeq5\times10^{18}\,$eV.}\\

\begin{figure}[t]
\begin{center}
\includegraphics[width=0.95\textwidth]{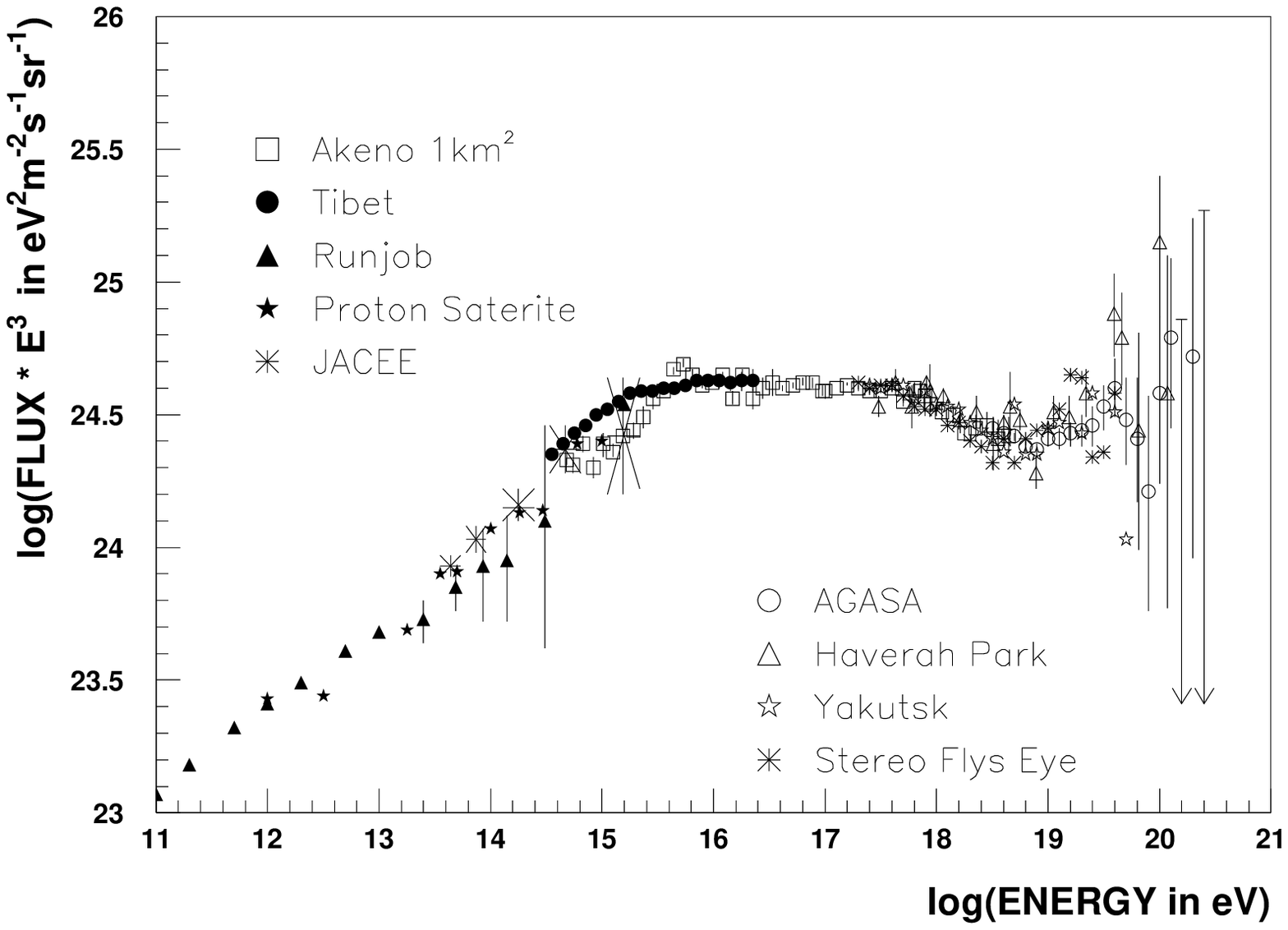}
\end{center}
\caption[...]{}
\label{fig1}
\end{figure}

Fig.~2: {Same as Fig.~\ref{fig1}, but focusing on the
high energy end above $10^{17}\,$eV (from Ref~\cite{data1}
with permission). The ``ankle'' is again visible at
$E\simeq5\times10^{18}\,$eV.}\\

Fig.~3: {The UHECR distribution of arrival times and energies (top),
the sky averaged spectrum (middle, with 1 sigma error bars showing
combined data from the Haverah Park~\cite{haverah}, the Fly's Eye~\cite{fe},
and the AGASA~\cite{agasa} experiments above $10^{19}\,$eV),
and the sky distribution in Galactic coordinates (bottom, with color
scale showing the intensity per solid angle) in the bottom-up
scenario with sources in the local Supercluster of galaxies
explained in the text. 20000 proton trajectories for 4
magnetic field realizations each were calculated. The cross-over
from the diffusive regime below $\simeq2\times10^{20}\,$eV to
the regime of rectilinear propagation at the highest energies
is clearly visible in the two upper panels.}\\

\begin{figure}[t]
\begin{center}
\includegraphics[height=0.95\textheight]{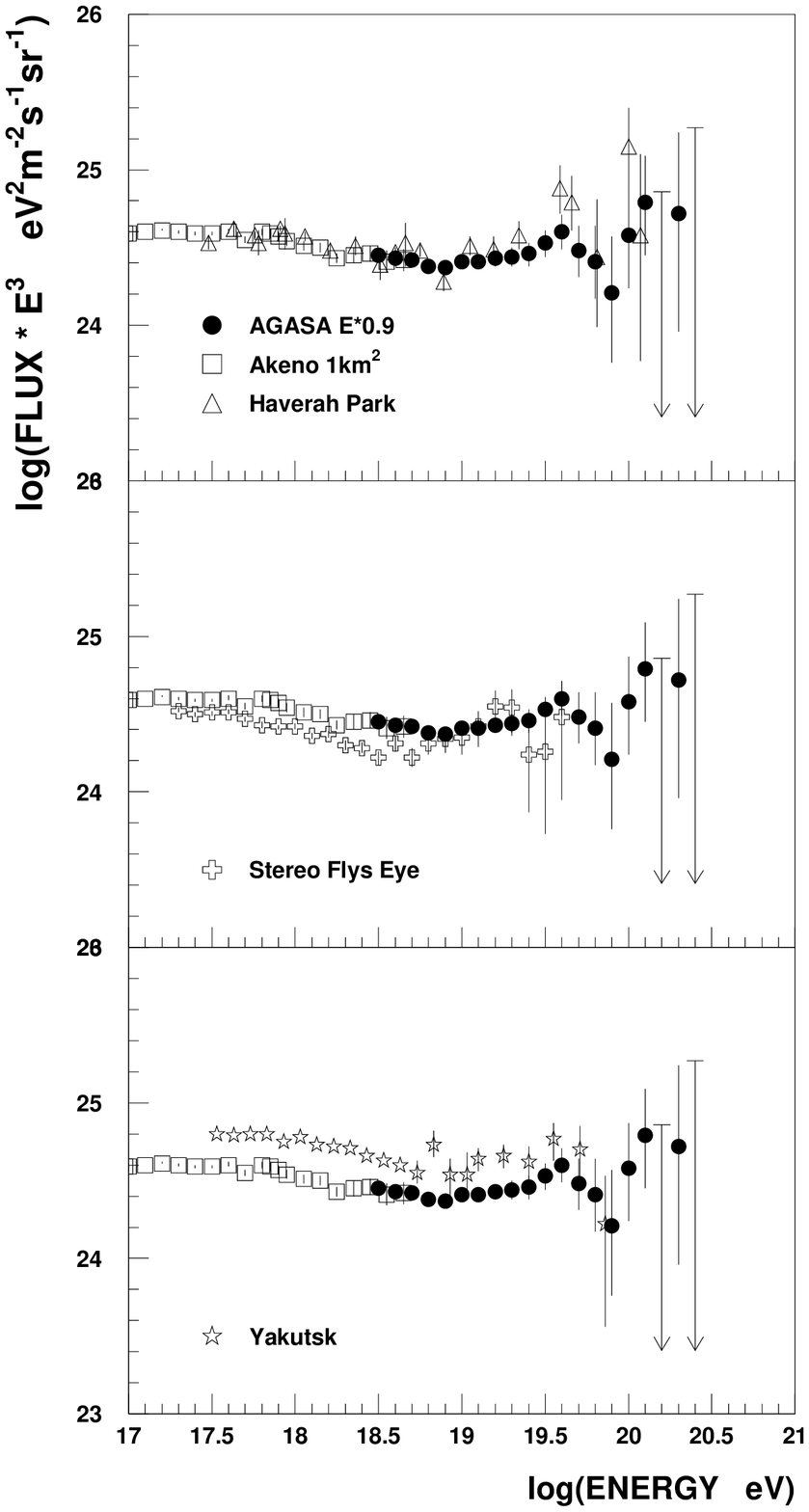}
\end{center}
\caption[...]{}
\label{fig2}
\end{figure}

\begin{figure}[t]
\begin{center}
\includegraphics[height=0.95\textheight]{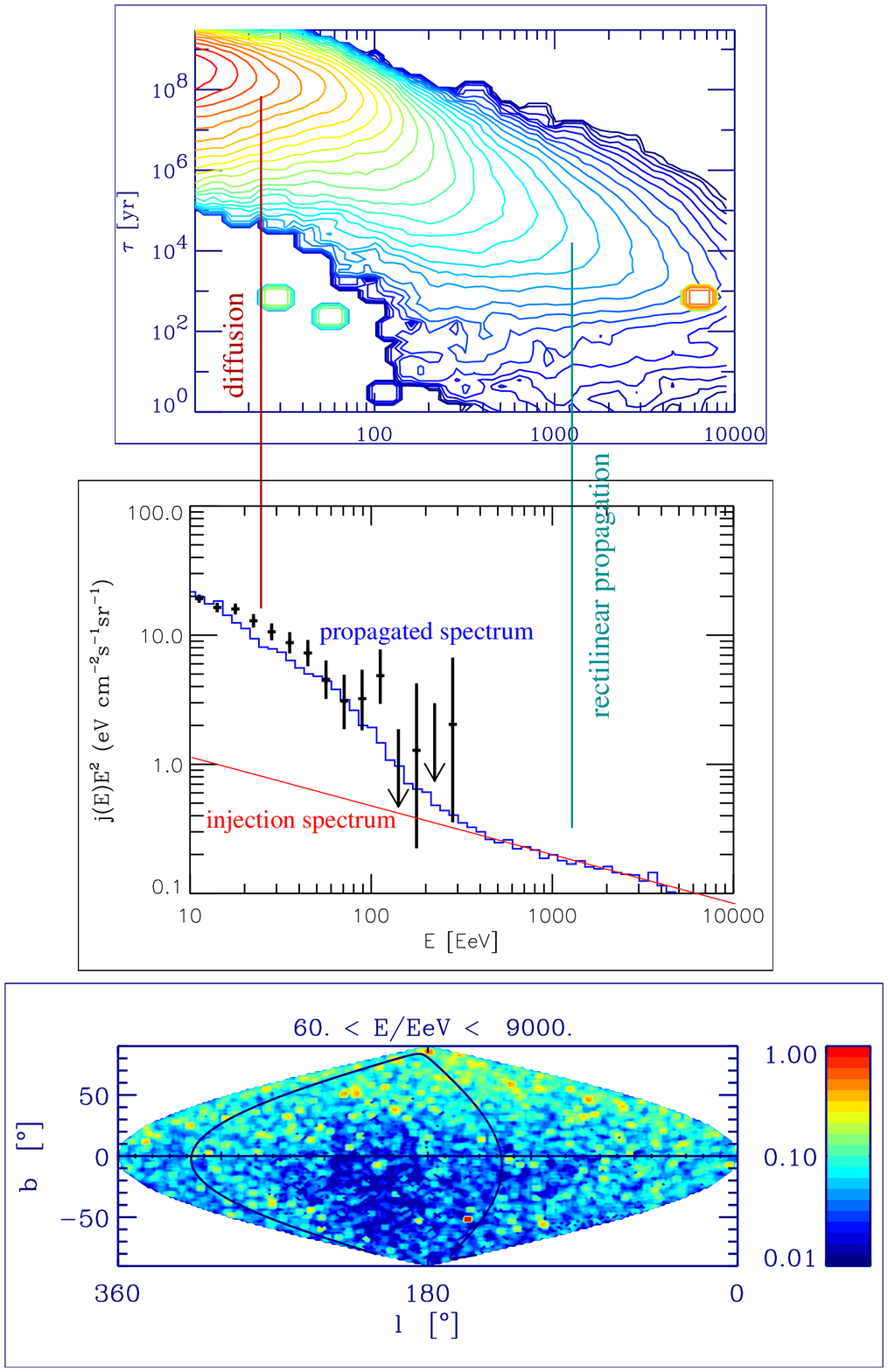}
\end{center}
\caption[...]{}
\label{fig3}
\end{figure}

\begin{figure}[t]
\begin{center}
\includegraphics[height=0.95\textheight]{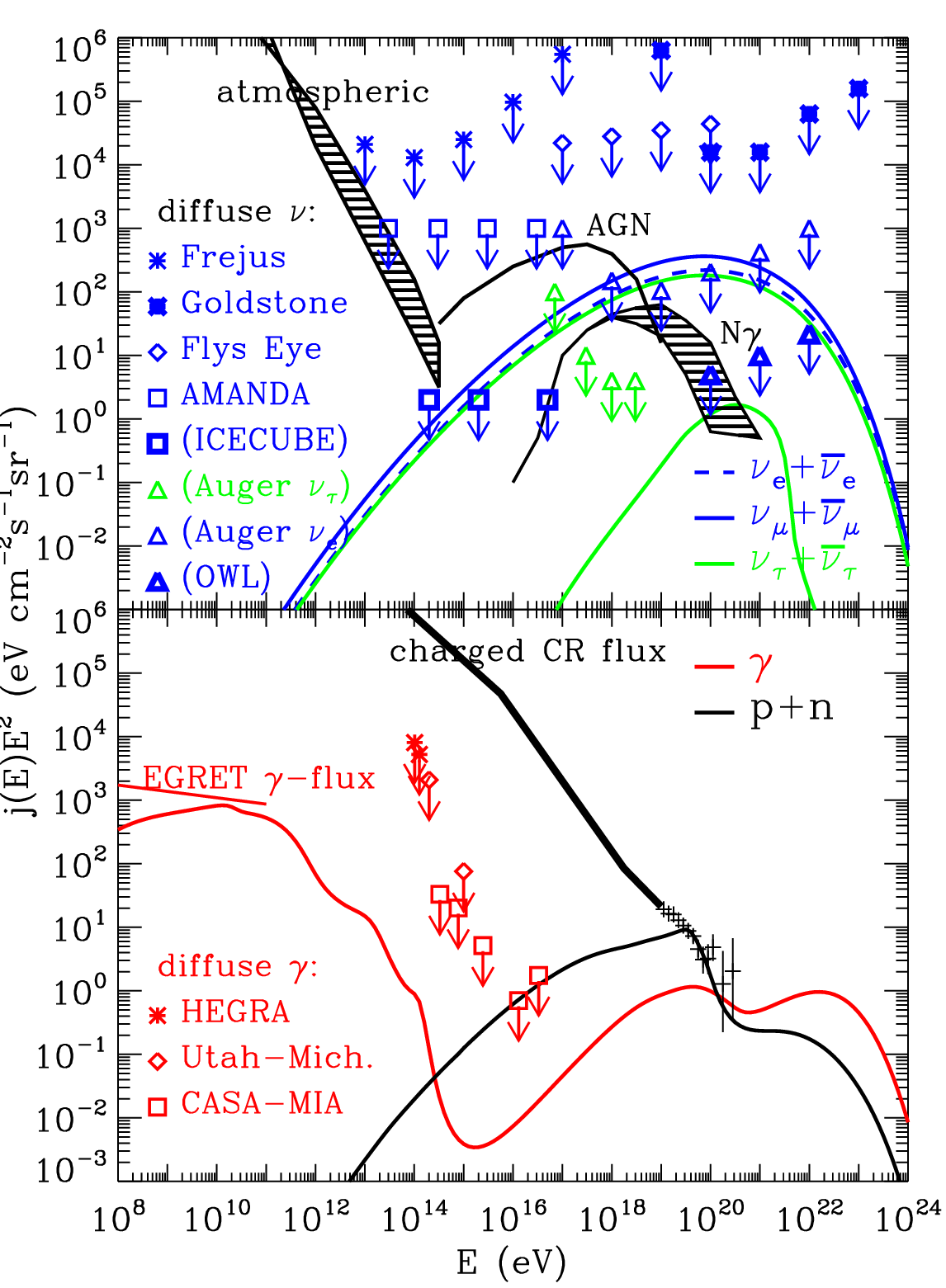}
\end{center}
\caption[...]{}
\label{fig4}
\end{figure}

Fig.~4: {All particle spectra for a top-down model involving the
decay into two quarks of non-relativistic X particles of mass $10^{16}$
GeV, released from homogeneously distributed topological defects. Lower
panel: The fluxes of the ``visible'' particles, nucleons
and $\gamma-$rays. 1 sigma error bars are as in Fig.~\ref{fig3} (see also
Fig.~\ref{fig2}). Also shown are piecewise power law fits to the
observed charged CR flux below $10^{19}\,$eV, the measurement
of the diffuse $\gamma$-ray flux between 30 MeV and 100 GeV by the
EGRET instrument~\cite{egret}, as well as upper limits on the diffuse
$\gamma-$ray flux from various experiments at higher energies
(see Ref.~\cite{bs} for more details). Upper panel: Neutrino fluxes. Shown
are experimental neutrino flux limits from the Fly's
Eye~\cite{baltrusaitis} and other experiments as indicated (see
Ref.~\cite{bs} for details), as well as projected neutrino
sensitivities of the Pierre Auger Project~\cite{has} (for electron
and tau neutrinos separately) and the proposed
space based OWL~\cite{owl} concept. For comparison also shown are
the atmospheric neutrino background (hatched region marked ``atmospheric''),
and neutrino flux predictions for a model of AGN optically thick to nucleons
(``AGN''), and for UHECR interactions with the CMB~\cite{pj}
(``$N\gamma$'', dashed range indicating typical uncertainties
for moderate source evolution).
The top-down fluxes are shown for electron-, muon, and tau-neutrinos
separately, assuming no (lower $\nu_\tau$-curve) and maximal
$\nu_\mu-\nu_\tau$ mixing (upper $\nu_\tau$-curve, which would then
equal the $\nu_\mu$-flux), respectively.}

\newpage

\newpage

\newpage

\newpage

\begin{thebibliography}{999}

\bibitem{hess} V.~F.~Hess, {\it Phys.~Z.} {\bf 13}, 1084 (1912).

\bibitem{auger_disc} P.~Auger, R.~Maze, T.~Grivet-Meyer,
{\it Acad\'emie des Sciences} {\bf 206}, 1721 (1938); P.~Auger, R.~Maze,
{\it ibid.} {\bf 207}, 228 (1938).

\bibitem{crbook} for a general introduction on cosmic rays
see, e.g., V.~S.~Berezinsky, S.~V.~Bulanov, V.~A.~Dogiel,
V.~L.~Ginzburg, V.~S.~Ptuskin, {\it Astrophysics of Cosmic
Rays} (North-Holland, Amsterdam, 1990); T.~K.~Gaisser, {\it
Cosmic Rays and Particle Physics}, Cambridge University Press
(Cambridge, 1998).

\bibitem{data1} M.~Nagano, A.~A.~Watson,
{\it Rev.~Mod.~Phys.} {\bf 72}, 689 (2000), and references therein.

\bibitem{data} for a summary of the data situation and experimental
issues see, e.g., S.~Yoshida, H.~Dai, {\it J.~Phys.~G} {\bf 24}, 905
(1998); X.~Bertou, M.~Boratav, A.~Letessier-Selvon,
{\it Int.~J.~Mod.~Phys.} {\bf A15}, 2181 (2000), and references therein.

\bibitem{bs} see, e.g., P.~Bhattacharjee, G.~Sigl,
{\it Phys.~Rep.} {\bf 327}, 109 (2000), and references therein.

\bibitem{reviews} see, e.g., J.~W.~Cronin, {\it Rev.~Mod.~Phys.}
{\bf 71}, S165 (1999); A.~V.~Olinto, {\it Phys.~Rept.} {\bf 333-334},
329 (2000); X.~Bertou, M.~Boratav, A.~Letessier-Selvon,
{\it Int.~J.~Mod.~Phys.} {\bf A15}, 2181 (2000).

\bibitem{gzk} K.~Greisen, {\it Phys.~Rev.~Lett.} {\bf 16}, 748 (1966);
G.~T.~Zatsepin, V.~A.~Kuzmin, {\it Pis'ma~Zh.~Eksp.~Teor.~Fiz.}
{\bf 4}, 114 (1966) [{\it JETP.~Lett.} {\bf 4}, 78 (1966)].

\bibitem{zburst1} T.~J.~Weiler, {\it Phys.~Rev.~Lett.} {\bf 49}, 234
(1982); {\it Astrophys.~J.} {\bf 285}, 495 (1984); {\it Astropart.~Phys.}
{\bf 11}, 317 (1999).

\bibitem{zburst2} S.~Yoshida, G.~Sigl, S.~Lee, {\it Phys.~Rev.~Lett.}
{\bf 81}, 5505 (1998); J.~J.~Blanco-Pillado, R.~A.~V\'{a}zquez,
E.~Zas, {\it Phys.~Rev.~D} {\bf 61}, 123003 (2000).

\bibitem{acc} A.~M.~Hillas, {\it Ann.~Rev.~Astron.~Astrophys.} {\bf 22},
425 (1984); C.~A.~Norman, D.~B.~Melrose, A.~Achterberg, {\it Astrophys.~J.}
{\bf 454}, 60 (1995).

\bibitem{beo} P.~Blasi, R.~I.~Epstein, A.~V.~Olinto,
{\it Astrophys.~J.~Lett.} {\bf 533}, L123 (2000).

\bibitem{dp} A.~Dar, R.~Plaga, {\it Astron.~Astrophys.} {\bf 349}, 259
(1999).

\bibitem{bier-rev} see, e.g., P.~L.~Biermann, {\it J.~Phys.~G:
Nucl.~Part.~Phys.} {\bf 23}, 1 (1997).

\bibitem{shocks} H.~Kang, J.~P.~Rachen, and P.~L.~Biermann,
{\it Mon.~Not.~R.~Soc.~Astron.} {\bf 286}, 257 (1997).

\bibitem{radio} P.~L.~Biermann, P.~A.~Strittmatter, {\it Astrophys.~J.}
{\bf 322}, 643 (1987).

\bibitem{stecker} F.~Stecker, {\it Astropart.~Phys.} {\bf 14}, 207
(2000).

\bibitem{waxman} E.~Waxman, {\it Phys.~Scripta} {\bf T85}, 117 (2000),
and references therein.

\bibitem{agasa_clu} N.~Hayashida et al., {\it Phys.~Rev.~Lett.} {\bf 77},
1000 (1996); M.~Takeda et al., {\it Astrophys.~J.} {\bf 522}, 225 (1999);
N.~Hayashida {\it et al.}, e-print astro-ph/0008102.

\bibitem{sources} G.~Sigl, D.~N.~Schramm, P.~Bhattacharjee,
{\it Astropart.~Phys.} {\bf 2}, 401 (1994); J.~W.~Elbert, P.~Sommers,
{\it Astrophys.~J.} {\bf 441}, 151 (1995).

\bibitem{strucmag} K.~T.~Kim et al., {\it Nature} {\bf 341}, 720 (1989);
T.~Clark, P.~P.~Kronberg, H.~B\"ohringer, submitted to
{\it Astrophys.~J.~Lett.}

\bibitem{mag} J.~P.~Vallee, {\it Fund.~Cosm.~Phys.} {\bf 19}, 1 (1997);
P.~Blasi, S.~Burles, A.~V.~Olinto, {\it Astrophys.~J.} {\bf 514}, L79
(1999).

\bibitem{highmag1} G.~Medina Tanco, {\it Astrophys.~J.~Lett.} {\bf 505},
L79 (1998); T.~Ensslin, e-print astro-ph/9906212;
E.-J.~Ahn, G.~Medina-Tanco, P.~L.~Biermann, T.~Stanev, e-print
astro-ph/9911123; G.~R.~Farrar, T.~Piran, {\it Phys.~Rev.~Lett.} {\bf 84},
3257 (2000); T.~Stanev, R.~Engel, A.~Muecke, R.~J.~Protheroe,
J.~P.~Rachen, {\it Phys.~Rev.~D} {\bf 62}, 093005 (2000).

\bibitem{highmag2} M.~Lemoine, G.~Sigl, P.~Biermann, e-print
astro-ph/9903124.

\bibitem{highmag3} G.~Sigl, M.~Lemoine, P.~Biermann, {\it Astropart.~Phys.}
{\bf 10}, 141 (1999); P.~Blasi, A.~V.~Olinto, {\it Phys.~Rev.~D.} {\bf 59},
023001 (1999).

\bibitem{deflec} G.~Sigl, M.~Lemoine, {\it Astropart.~Phys.} {\bf 9},
65 (1998).

\bibitem{gr} see, e.g., D.~Grasso, H.~R.~Rubinstein, e-print
astro-ph/0009061, to appear in {\it Phys.~Rept.}

\bibitem{kt} see, e.g., E.~W.~Kolb, M.~S.~Turner,
{\it The Early Universe} (Addison-Wesley, Redwood City, California, 1990). 

\bibitem{agasa} N.~Hayashida et al., {\it Phys.~Rev.~Lett.} {\bf 73},
3491 (1994); S.~Yoshida et al., {\it Astropart.~Phys.} {\bf 3}, 105 (1995);
M.~Takeda {\it et al.}, {\it Phys.~Rev.~Lett.} {\bf 81}, 1163 (1998); see
also {\sf http://icrsun.icrr.u-tokyo.ac.jp/as/project/agasa.html}.

\bibitem{fe} D.~J.~Bird et al., {\it Phys.~Rev.~Lett.} {\bf 71}, 3401
(1993); {\it Astrophys.~J.} {\bf 424}, 491 (1994); {\it ibid.} {\bf 441},
144 (1995).

\bibitem{hires} S.~C.~Corbat\'{o} {\it et al.}, {\it Nucl.~Phys.~B
(Proc. Suppl.)} {\bf 28B}, 36 (1992); see also
{\sf http://hires.physics.utah.edu/}.

\bibitem{tel_array} M.~Teshima et al., {\it Nucl.~Phys.~B (Proc.~Suppl.)}
{\bf 28B}, 169 (1992); see also {\sf http://www-ta.icrr.u-tokyo.ac.jp/}.

\bibitem{auger} J.~W.~Cronin, {\it Nucl.~Phys.~B (Proc.~Suppl.)} {\bf 28B},
213 (1992); The Pierre Auger Observatory Design Report (2nd
edition), March 1997; see also
{\sf http://http://www.auger.org/} and
{\sf http://www-lpnhep.in2p3.fr/auger/welcome.html}.

\bibitem{owl} D.~B.~Cline, F.~W.~Stecker, OWL/AirWatch science
white paper, e-print astro-ph/0003459;
see also {\sf http://lheawww.gsfc.nasa.gov/docs/gamcosray/hecr/OWL/}.

\bibitem{euso} See {\sf http://www.ifcai.pa.cnr.it/Ifcai/euso.html}.

\bibitem{has} J.~J.~Blanco-Pillado, R.~A.~V\'{a}zquez,
E.~Zas, {\it Phys.~Rev.~Lett.} {\bf 78}, 3614 (1997); K.~S.~Capelle,
J.~W.~Cronin, G.~Parente, E.~Zas, {\it Astropart.~Phys.} {\bf 8}, 321
(1998); A.~Letessier-Selvon, e-print astro-ph/0009444.

\bibitem{gut} see, e.g., S.~Weinberg, {\it The Quantum Theory of
Fields Vol 2: Modern Applications}, Cambridge University Press
(Cambridge, 1996).

\bibitem{vachaspati} see, e.g., T.~Vachaspati, {\it Contemp.~Phys.}
{\bf 39}, 225 (1998).

\bibitem{kuz-tak} for a brief review see V.~Kuzmin, I.~Tkachev,
{\it Phys.~Rept.} {\bf 320}, 199 (1999).

\bibitem{egret} P.~Sreekumar {\it et al.}, {\it Astrophys.~J.} {\bf 494}, 523
(1998).

\bibitem{haverah} See, e.g., M.~A.~Lawrence, R.~J.~O.~Reid,
A.~A.~Watson, {\it J.~Phys.~G Nucl.~Part.~Phys.} {\bf 17}, 733 (1991), and
references therein; see also
{\sf http://ast.leeds.ac.uk/haverah/hav-home.html}.

\bibitem{slby} G.~Sigl, S.~Lee, P.~Bhattacharjee, S.~Yoshida, 
{\it Phys.~Rev.~D} {\bf 59}, 043504 (1999).

\bibitem{wbbound} E.~Waxman, J.~Bahcall, {\it Phys.~Rev.~D.} {\bf 59},
023002 (1999); J.~Bahcall, E.~Waxman, e-print hep-ph/9902383;
K.~Mannheim, R.~J.~Protheroe, J.~P.~Rachen,
e-print astro-ph/9812398, to appear in {\it Phys.~Rev.~D.}; J.~P.~Rachen,
R.~J.~Protheroe, K.~Mannheim, e-print astro-ph/9908031.

\bibitem{gamma} F.~Halzen, R.~A.~V'{a}zques, T.~Stanev,
H.~P.~Vankov, {\it Astropart.~Phys.} {\bf 3}, 151 (1995); M.~Ave,
J.~A.~Hinton, R.~A.~V'{a}zques, A.~A.~Watson, E.~Zas,
{\it Phys.~Rev.~Lett.} {\bf 85} 2244 (2000).

\bibitem{cfk} G.~R.~Farrar, {\it Phys.~Rev.~Lett.} {\bf 76}, 4111 (1996);
D.~J.~H.~Chung, G.~R.~Farrar, E.~W.~Kolb, {\it Phys.~Rev.~D}
{\bf 57}, 4696 (1998).

\bibitem{gluino} I.~F.~Albuquerque {\it et al.} (E761 collaboration),
{\it Phys.~Rev.~Lett.} {\bf 78}, 3252 (1997); A.~Alavi-Harati {\it et al.}
(KTeV collaboration), {\it Phys.~Rev.~Lett.} {\bf 83}, 2128 (1999).

\bibitem{monopole} see S.~D.~Wick, T.~W.~Kephart, T.~J.~Weiler,
P.~L.~Biermann, e-print astro-ph/0001233, and references
therein.

\bibitem{bp} S.~Bonazzola, P.~Peter, {\it Astropart.~Phys}. {\bf 7}, 161
(1997).

\bibitem{corr} G.~R.~Farrar, P.~L.~Biermann, {\it Phys.~Rev.~Lett.} {\bf 81},
3579 (1998); C.~M.~Hoffman, {\it ibid.} {\bf 83}, 2471 (1999); G.~R.~Farrar,
P.~L.~Biermann, {\it ibid.} {\bf 83}, 2472 (1999); G.~Sigl, D.~F.~Torres,
L.~A.~Anchordoqui, G.~E.~Romero, e-print astro-ph/0008363;
A.~Virmani {\it et al.}, e-print astro-ph/0010235.

\bibitem{gw} H.~Goldberg and T.~J.~Weiler, {\it Phys.~Rev.~D} {\bf 59},
113005 (1999).

\bibitem{kovesi-domokos} G.~Domokos, S.~Kovesi-Domokos,
{\it Phys.~Rev.~Lett.} {\bf 82}, 1366 (1999).

\bibitem{tev-qg} see, e.g., N.~Arkani-Hamed, S.~Dimopoulos, G.~Dvali,
{\it Phys.~Rev.~D} {\bf 59}, 086004 (1999).

\bibitem{nucross} S.~Nussinov, R.~Shrock, {\it Phys.~Rev.~D} {\bf 59},
105002 (1999); P.~Jain, D.~W.~McKay, S.~Panda, J.~P.~Ralston,
{\it Phys. Lett.} {\bf B484}, 267 (2000); J.~P.~Ralston, P.~Jain,
D.~W.~McKay, S.~Panda, e-print hep-ph/0008153.

\bibitem{cpp} see, e.g., S.~Cullen, M.~Perelstein,
M.~E.~Peskin, {\it Phys.~Rev.D} {\bf 62}, 055012 (2000), and references
therein.

\bibitem{sn87a} S.~Cullen, M.~Perelstein, {\it Phys.~Rev.~Lett.}
{\bf 83}, 268 (1999); V.~Barger, T.~Han, C.~Kao, R.-J.~Zhang,
{\it Phys.~Lett.~B} {\bf 461}, 34 (1999).

\bibitem{tos} C.~Tyler, A.~Olinto, G.~Sigl, e-print hep-ph/0002257.

\bibitem{lorentz} see, e.g., S.~Coleman, S.~L.~Glashow, {\it Nucl.~Phys.}
{\bf B574}, 130 (2000); L.~Gonzalez-Mestres, {\it Nucl.~Phys.~B (Proc.~Suppl.)}
{\bf 48}, 131 (1996), and references therein.

\bibitem{abgg} see, e.g., R.~Aloisio, P.~Blasi, P.~L.~Ghia,
A.~F.~Grillo, {\it Phys.~Rev.~D} {\bf 62}, 053010 (2000).

\bibitem{baltrusaitis} R.~M.~Baltrusaitis {\it et al.},
{\it Astrophys.~J.} {\bf 281}, L9 (1984); {\it Phys.~Rev.~D} {\bf 31},
2192 (1985).

\bibitem{pj} R.~J.~Protheroe, P.~A.~Johnson,
{\it Astropart.~Phys.} {\bf 4}, 253 (1996), and erratum {\it ibid.}
{\bf 5}, 215 (1996).

\bibitem{ack} I acknowledge P.~Biermann, P.~Blasi, M.~Boratav,
T.~Ensslin, P.~Peter, R.~Plaga, and G.~Raffelt for very helpful comments
on the manuscript.

\end{thebibliography}
\end{document}